\documentclass[twocolumn,superscriptaddress,showpacs,oneside,prl]{revtex4}%
\usepackage{amsmath}
\usepackage{amssymb}
\usepackage{amsfonts}
\usepackage{graphicx}%
\begin{document}
\title{Experimental Test of the Kochen-Specker Theorem for Single Qubits using
Positive Operator-Valued Measures}
\author{Qiang Zhang}
\affiliation{Hefei National Laboratory for Physical Sciences at Microscale \& Department of
Modern Physics, University of Science and Technology of China, Hefei, Anhui
230026, P.R. China}
\author{Hui Li}
\affiliation{Hefei National Laboratory for Physical Sciences at Microscale \& Department of
Modern Physics, University of Science and Technology of China, Hefei, Anhui
230026, P.R. China}
\author{Tao Yang}
\affiliation{Hefei National Laboratory for Physical Sciences at Microscale \& Department of
Modern Physics, University of Science and Technology of China, Hefei, Anhui
230026, P.R. China}
\author{Juan Yin}
\affiliation{Hefei National Laboratory for Physical Sciences at Microscale \& Department of
Modern Physics, University of Science and Technology of China, Hefei, Anhui
230026, P.R. China}
\author{Jiangfeng Du}
\affiliation{Hefei National Laboratory for Physical Sciences at
Microscale \& Department of Modern Physics, University of Science
and Technology of China, Hefei, Anhui 230026, P.R. China}
\affiliation{Department of Phsics,National University of
Singapore, 2 Science Driver 3, Singapore, 117542}
\author{Jian-Wei Pan}
\affiliation{Hefei National Laboratory for Physical Sciences at
Microscale \& Department of Modern Physics, University of Science
and Technology of China, Hefei, Anhui 230026, P.R. China}
\affiliation{Physikaliches Institut, Universit\"{a}t Heidelberg,
Philisophenweg 12, 69120 Heidelberg, Germany}
\begin{abstract}
We present an experimental scheme for the implementation of
arbitrary generalized measurements, represented by
positive-operator valued measures, on the polarization of single
photons, using linear optical devices. Further, we experimentally
test a Kochen-Specker theorem for single qubits using positive
operator-valued measures. Our experimental results for the first
time disprove non-contextual hidden-variable theories, even for
single qubits.

\end{abstract}
\pacs{03.65.Ta, 03.65.Ud, 42.50.Xa}
\maketitle

Hidden-variable theories (HV), inspired by Einstein, Podolsky, and
Rosen (EPR) with their famous paradox \cite{r1}, has attracted
broad interests. In 1960's, Bell published his famous inequality
\cite{r2} that revealed the quantitative contradiction between
local hidden-variable (LHV) theories and quantum mechanics (QM),
leading to experimental tests on this fundamental problem. A
number of experiments \cite{r3} have observed the
incompatibilities of LHV theories and experimental data,
confirming that only by QM can the experimental results be
correctly explained. There is another type, in fact a general
type, of hidden-variable theories, i.e. the noncontextual
hidden-variable (NCHV) theories. In such theories, values of
physical observables are the same whatever the experimental
context in which they appear. Kochen-Specker (KS) theorem
\cite{r4} dictates the contradiction between such NCHV theories
and QM. Recently, an all-or-nothing--type Kochen-Specker theorem
has been experimentally tested \cite{r5}. Traditional KS theorem
applies only to physical systems described by Hilbert spaces of
dimension three or higher. However, it has been proved that KS
theorems can be proved for a single two-level system (a qubit)
\cite{cabello}, using generalized measurements represented by
positive operator-valued measures (POVMs) \cite{book1,book2},
which have found applications in various fields of physical
research \cite{r6,cabello}.

In this paper, we present an experimental scheme for
implementation of arbitrary generalized measurements on
polarization states of single photons, using linear optical
devices. One interesting and important application is to
experimentally test the Kochen-Specker (KS) theorem for a single
qubit using POVMs \cite{cabello}, as will be presented in this
paper. We believe this is the first experimental test of a KS
theorem for single qubits. Our results show that even for a single
qubit NCHV theories cannot be consistent with experiments.

The POVM elements can always be expressed as linear combinations
of one-dimensional operators, each of that has one and only one
non-zero eigenvalue. Therefore it is sufficient \cite{davies} to
consider POVMs that consist all of one-dimensional operators.
Based on the Neumark's theorem \cite{neumark}, it can be proved
that either a $2\mathcal{N}$- or a $(2\mathcal{N}-1)$-element POVM
in $\mathbb{C}^{2}$\ can be realized via some orthogonal
measurement (OM) in a $2\mathcal{N}$-dimensional Hilbert space $\mathbb{C}^{\mathcal{N}%
}\otimes\mathbb{C}^{2}$. First, we consider the POVM associated with the
$2\mathcal{N}$-element set $\{E_{d}\}$ ($d=1,\cdots,2\mathcal{N}$) of the form%
\begin{equation}
E_{d}=|\widetilde{\psi}_{d}\rangle\langle\widetilde{\psi}_{d}|,\label{eq 9}%
\end{equation}
where $|\widetilde{\psi}_{d}\rangle\in\mathbb{C}^{2}$ (not normalized) and
$\sum_{d}E_{d}=I$ (the identity), there always exist vectors $|\widetilde
{\phi}_{d}\rangle$ such that the following vectors%
\begin{equation}
|\varphi_{d}\rangle=\left(
\begin{array}
[c]{c}%
|\widetilde{\psi}_{d}\rangle\\
|\widetilde{\phi}_{d}\rangle
\end{array}
\right)  \in\mathbb{C}^{\mathcal{N}}\otimes\mathbb{C}^{2}\label{eq 10}%
\end{equation}
are orthonormal. The set
$\{|\varphi_{d}\rangle\langle\varphi_{d}|\}$ thus represents the
OM in $\mathbb{C}^{\mathcal{N}}\otimes\mathbb{C}^{2}$ that
realizes the POVM $\{E_{d}\}$ in $\mathbb{C}^{2}$. The POVM
$\{E_{d}\}$ on the
state%
\begin{equation}
|\Psi\rangle=\left(
\begin{array}
[c]{c}%
\alpha\\
\beta
\end{array}
\right)  \in\mathbb{C}^{2}\ \left(  |\alpha|^{2}+|\beta|^{2}=1\right)
\label{eq st}%
\end{equation}
can then be realized via the OM
$\{|\varphi_{d}\rangle\langle\varphi_{d}|\}$
on the state%
\begin{equation}
|\Phi\rangle=\left(  \alpha,\beta,0,\cdots,0\right)  ^{Transpose}\in
\mathbb{C}^{\mathcal{N}}\otimes\mathbb{C}^{2}.\label{eq st'}%
\end{equation}

Now consider the POVM associated with a $(2\mathcal{N}-1)$-element
set $\{E_{d}^{\prime}\}$ ($d=1,\cdots,2\mathcal{N}-1$) of the form
similar to Eq.
(\ref{eq 9}):%
\begin{equation}
E_{d}^{\prime}=|\widetilde{\psi}_{d}^{\prime}\rangle\langle\widetilde{\psi
}_{d}^{\prime}|.\label{eq 9'}%
\end{equation}
There also exist vectors $|\widetilde{\phi}_{d}^{\prime}\rangle$ such that the
following $2\mathcal{N}$ vectors%
\begin{equation}
|\varphi_{d}\rangle=\left(
\begin{array}
[c]{c}%
|\widetilde{\psi}_{d}^{\prime}\rangle\\
|\widetilde{\phi}_{d}^{\prime}\rangle\\
0
\end{array}
\right)  ,\qquad|\varphi_{2\mathcal{N}}\rangle=\left(
\begin{array}
[c]{c}%
0\\
\vdots\\
0\\
1
\end{array}
\right)  \in\mathbb{C}^{\mathcal{N}}\otimes\mathbb{C}^{2},\label{eq 10'}%
\end{equation}
are orthonormal. Hence the POVM $\{E_{d}^{\prime}\}$ on the state
$|\Psi\rangle$ in Eq. (\ref{eq st}) could be realized via the OM
$\{|\varphi_{d}\rangle\langle\varphi_{d}|,|\varphi_{2\mathcal{N}}%
\rangle\langle\varphi_{2\mathcal{N}}|\}$ on the state $|\Phi\rangle$ in Eq.
(\ref{eq st'}). Here we shall note that since $\langle\varphi_{2\mathcal{N}%
}|\Phi\rangle=0$, the projector $|\varphi_{2\mathcal{N}}\rangle\langle
\varphi_{2\mathcal{N}}|$ will always yield null outcome when measuring the
state $|\Phi\rangle$. Only the projectors $\{|\varphi_{d}\rangle\langle
\varphi_{d}|\}$ could yield non-null outcomes, precisely corresponding to the
POVM $\{E_{d}^{\prime}\}$.

For POVMs on the polarization states of a single photon, $\mathcal{N}$
different paths could be used to span the ancilla Hilbert space $\mathbb{C}%
^{\mathcal{N}}$. We denote them by mode states $|k\rangle$ ($k=1,\cdots
,\mathcal{N}$). States $|\Psi\rangle$ in Eq. (\ref{eq st}) and $|\Phi\rangle$
in Eq. (\ref{eq st'}) could thus be written as%
\begin{align}
|\Psi\rangle &  =\alpha|\text{H}\rangle+\beta|\text{V}\rangle,\\
|\Phi\rangle &  =|1\rangle\otimes\left(  \alpha|\text{H}\rangle+\beta
|\text{V}\rangle\right)  .
\end{align}
where $|\text{H}\rangle$ ($|\text{V}\rangle$) denotes the
horizontal (vertical) polarization. The crucial step is then to
perform the OM given in Eq. (\ref{eq 10}) or (\ref{eq 10'}) on
$|\Phi\rangle$.

Indeed, one can always find a $2\mathcal{N}$-dimensional unitary operator, say
$U_{2\mathcal{N}}$, that fulfills exactly the following transformation,%
\begin{equation}
|\varphi_{2k-1}\rangle\ ^{\underrightarrow{\ U_{2\mathcal{N}}\ }}%
\ |k,\text{H}\rangle,\qquad|\varphi_{2k}\rangle\ ^{\underrightarrow
{\ U_{2\mathcal{N}}\ }}\ |k,\text{V}\rangle,\label{eq 14}%
\end{equation}
with $k=1,\cdots,\mathcal{N}$ and
$|k,\text{H}\rangle=|k\rangle\otimes |\text{H}\rangle$ etc.
Therefore the OM on the state $|1\rangle\otimes\left(
\alpha|\text{H}\rangle+\beta|\text{V}\rangle\right)  $, and
consequently POVM $\{E_{d}\}$ on
$\alpha|\text{H}\rangle+\beta|\text{V}\rangle$, can be realized
by performing OM $\{|k,\text{H}\rangle\langle k,\text{H}|,|k,\text{V}%
\rangle\langle k,\text{V}|\}$ on $U_{2\mathcal{N}}\left[  |1\rangle
\otimes\left(  a|\text{H}\rangle+b|\text{V}\rangle\right)  \right]  $. It is
obvious that OM  $\{|k,\text{H}\rangle\langle k,\text{H}|,|k,\text{V}%
\rangle\langle k,\text{V}|\}$ can be carried out by placing
polarizing beam splitters (PBS) followed by single-photon
detectors at out-ports of every path. In what follows, we describe
the scheme for
implementing arbitrary unitary operators on the Hilbert space $\mathbb{C}%
^{\mathcal{N}}\otimes\mathbb{C}^{2}$, of which $\mathbb{C}^{\mathcal{N}}$ is
spanned by paths while $\mathbb{C}^{2}$\ by polarization.

\begin{figure}[ptb]
\includegraphics[width=.9\columnwidth]{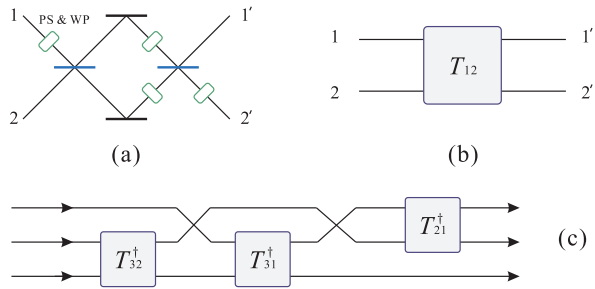}\caption{(a) A Mach-Zehnder
interferometer with properly placed phase shifters and wave plates
(PS\&WP's) can be used as the basic building block of any
$2N$-dimensional unitary matrix. The Mach-Zehnder interferometer
can be represented by the abstract four-port device in (b). (c)
Three Mach-Zehnder interferometer devices
$T_{pq}$ are enough to build any unitary operators on $\mathbb{C}^{3}%
\otimes\mathbb{C}^{2}$.}%
\label{Fig1}%
\end{figure}

The technique employed here is similar as in Ref. \cite{reck}. As
shown in Ref. \cite{eng}, the most general element of $U\left(
4\right)  $ can be realized by a Mach-Zehnder (MZ) interferometer
with four specific unitary operation on the polarization, which
could be realized by a proper combination of phase shifters,
quarter- and half-wave plates [see in Fig. \ref{Fig1}(a)]. The
most crucial observation is that an arbitrary
$2\mathcal{N}$-dimensional unitary operator can be factorized into
a product of block matrices which can be realized by a $U\left(
4\right)  $ operation on the Hilbert space spanned by the
polarizations and two different paths.

We define a matrix $T_{pq}$ ($p,q=1,\cdots,\mathcal{N}$) which is a
$2\mathcal{N}$-dimensional identity matrix with the elements $I_{ij}$
($i=2p-1,2p;j=2q-1,2q$) replaced by corresponding elements of the $U\left(
4\right)  $ operator of a MZ as in Fig. \ref{Fig1}(a,b). Using methods similar
to Gaussian elimination, by being multiplied from the right with a succession
of MZ $T_{\mathcal{N},q}$ ($q=\mathcal{N}-1,\cdots,1$), a $2\mathcal{N}%
$-dimensional unitary operator $U_{2\mathcal{N}}$ can be reduced into a direct
sum of a $\left(  2\mathcal{N}-2\right)  $-dimensional unitary operator
$U_{2\mathcal{N}-2}$ and the $2$-dimensional identity operator $I_{2}$:%
\begin{equation}
U_{2\mathcal{N}}\cdot T_{\mathcal{N},\mathcal{N}-1}\cdot T_{\mathcal{N}%
,\mathcal{N}-2}\cdot\cdots\cdot T_{\mathcal{N},1}=\left(
\begin{array}
[c]{cc}%
U_{2\mathcal{N}-2} & 0\\
0 & I_{2}%
\end{array}
\right)  .\label{eq 11}%
\end{equation}
The sequence of MZ transformations can be applied recursively to the matrix
with reduced dimensions. Thus we can make the resulting matrix equal to the
identity,%
\begin{equation}
U_{2\mathcal{N}}\cdot T_{\mathcal{N},\mathcal{N}-1}\cdot T_{\mathcal{N}%
,\mathcal{N}-2}\cdots\cdot T_{2,1}=I_{2\mathcal{N}},\label{eq 12}%
\end{equation}
and consequently we have%
\begin{align}
U_{2\mathcal{N}}  &  =\left(  T_{\mathcal{N},\mathcal{N}-1}\cdot
T_{\mathcal{N},\mathcal{N}-2}\cdot\cdots\cdot T_{2,1}\right)  ^{-1}\nonumber\\
&  =T_{2,1}^{\dag}\cdot\cdots\cdot T_{\mathcal{N},\mathcal{N}-2}^{\dag}\cdot
T_{\mathcal{N},\mathcal{N}-1}^{\dag}.\label{eq 13}%
\end{align}
Therefore the unitary transformation $U_{2\mathcal{N}}$\ could be realized by
recursively placing proper MZs shown in Fig. \ref{Fig1}(a). As an example, we
present in Fig. \ref{Fig1}(c) the setup for a general unitary matrix on
$\mathbb{C}^{3}\otimes\mathbb{C}^{2}$. We shall note that our scheme is
similar to the one proposed in Ref. \cite{reck}, where however the
polarization was not involved. Once all unitary transformations on
$\mathbb{C}^{\mathcal{N}}\otimes\mathbb{C}^{2}$ becomes realizable, it is
possible to perform any POVM on the polarization states of single photons. For
the task of performing a POVM on the polarization states of single photons,
the full setup according to Eq. (\ref{eq 13}) contains MZs of which the inputs
and outputs are exactly vacuum states. These MZs can be simply removed [e.g.
the $T_{32}^{\dag}$ in Fig. \ref{Fig1}(c)], and the setup of the POVM can
hence be further simplified.

One of the applications of the above scheme is to the optical test of a KS
theorem for a single qubit using positive operator-valued measures, proposed
by M. Nakamura (see Ref. [28] of \cite{cabello}). We now briefly describe the
KS theorem tested in this paper, which is in fact a simpler version of the one
proved in Ref. \cite{cabello}.

\begin{figure}[ptb]
\includegraphics[width=.9\columnwidth]{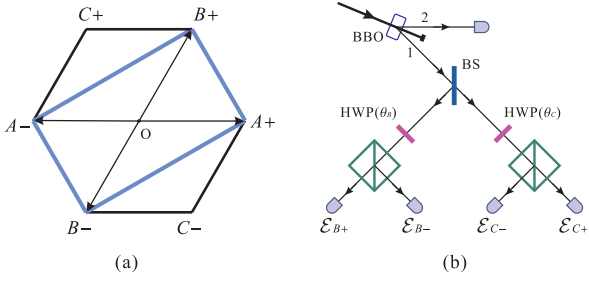}\caption{(a) Notation for the
six vertices of the regular hexagon: $A+$ is the antipode of $A-$, etc. $O$ is
the center of the regular hexagon. The rectangle formed by $A\pm$ and $B\pm$
is one of the three inscribed (sharing vertices) in the regular hexagon. It
corresponds to the four-element POVM $\{\mathcal{E}_{A\pm},\mathcal{E}_{B\pm
}\}$. (b) The schematic setup for the realization of the POVM $\{\mathcal{E}%
_{B\pm},\mathcal{E}_{C\pm}\}$. The beam splitter is a 50:50 one. The two
half-wave plates are set at $\theta_{B}=15{{}^{\circ}}$ and $\theta_{C}%
=30{{}^{\circ}}$.}%
\label{Fig2}%
\end{figure}

Let $A$, $B$, $C$ be the three directions obtained by joining the center of a
regular hexagon with its three non-antipodal vertices, as illustrated in Fig.
\ref{Fig2}(a). We can define six positive-semidefinite operators,
$\{\mathcal{E}_{A\pm},\mathcal{E}_{B\pm},\mathcal{E}_{C\pm}\}$, as follows.%
\begin{align}
\mathcal{E}_{A\pm}  &  =\frac{1}{2}\left\vert A\pm\right\rangle \left\langle
A\pm\right\vert ,\qquad\mathcal{E}_{B\pm}=\frac{1}{2}\left\vert B\pm
\right\rangle \left\langle B\pm\right\vert ,\nonumber\\
\mathcal{E}_{C\pm}  &  =\frac{1}{2}\left\vert C\pm\right\rangle \left\langle
C\pm\right\vert .\label{eq 1}%
\end{align}
These six operators can be used to construct three four-element POVMs:%
\begin{equation}
\{\mathcal{E}_{A\pm},\mathcal{E}_{B\pm}\},\qquad\{\mathcal{E}_{B\pm
},\mathcal{E}_{C\pm}\},\qquad\{\mathcal{E}_{C\pm},\mathcal{E}_{A\pm
}\}.\label{eq povm}%
\end{equation}
Geometrically, as shown in Fig. \ref{Fig2}(a), there are only three rectangles
share inscribed (sharing vertices) in a regular hexagon. All of them share the
same center, and any two rectangles share two antipodal vertices. Each
rectangle allows us to define a four-element POVM, which can be expressed as%
\begin{align}
\mathcal{E}_{A+}+\mathcal{E}_{B+}+\mathcal{E}_{A-}+\mathcal{E}_{B-}  &
=I_{2},\nonumber\\
\mathcal{E}_{B+}+\mathcal{E}_{C+}+\mathcal{E}_{B-}+\mathcal{E}_{C-}  &
=I_{2},\nonumber\\
\mathcal{E}_{C+}+\mathcal{E}_{A+}+\mathcal{E}_{C-}+\mathcal{E}_{A-}  &
=I_{2}.\label{eq 2}%
\end{align}
Each equation contains four positive-semidefinite operators, summing up to the
identity. A NCHV theory must assign the answer \textit{yes} to one and only
one of these four operators. However, such an assignment is impossible, since
each operator appears twice in Eqs. (\ref{eq 2}), so that the total number of
\textit{yes} answers must be an even number, while the number of possible
\textit{yes} answers, is three.

Experimentally, a qubit can be represented by polarization of a single photon.
In the basis of horizontal polarization $|\text{H}\rangle$ and vertical
polarization $\left\vert \text{V}\right\rangle $, we can write,%
\begin{equation}%
\begin{array}
[c]{ll}%
|A+\rangle=|\text{H}\rangle, & |A-\rangle=|\text{V}\rangle,\\
|B+\rangle=\frac{\sqrt{3}}{2}|\text{H}\rangle+\frac{1}{2}|\text{V}\rangle, &
|B-\rangle=\frac{1}{2}|\text{H}\rangle-\frac{\sqrt{3}}{2}|\text{V}\rangle,\\
|C+\rangle=\frac{1}{2}|\text{H}\rangle+\frac{\sqrt{3}}{2}|\text{V}\rangle, &
|C-\rangle=\frac{\sqrt{3}}{2}|\text{H}\rangle-\frac{1}{2}|\text{V}\rangle.
\end{array}
\label{eq 3}%
\end{equation}

Taking the implementation of the POVM
$\{\mathcal{E}_{B\pm},\mathcal{E}_{C\pm }\}$ as an example, the
corresponding OM can be constructed with the following four
orthonormal states in $\mathbb{C}^{2}\otimes\mathbb{C}^{2} $
[see Eq. (\ref{eq 10})]:%
\begin{align}
\frac{1}{\sqrt{2}}%
\begin{pmatrix}
|B\pm\rangle\\
i|B\pm\rangle
\end{pmatrix}
&  =\frac{|1\rangle+i|2\rangle}{\sqrt{2}}\otimes|B\pm\rangle,\nonumber\\
\frac{1}{\sqrt{2}}%
\begin{pmatrix}
|C\pm\rangle\\
-i|C\pm\rangle
\end{pmatrix}
&  =\frac{|1\rangle-i|2\rangle}{\sqrt{2}}\otimes|C\pm\rangle.\label{eq OM}%
\end{align}
Due to the specific form of Eqs. (\ref{eq OM}), the unitary
transformation shown in Eq. (\ref{eq 14}) can be realized simply
by a single 50:50 beam splitter (BS) followed by two unitary
transformations on the polarization states, without the necessity
of a full setup of the MZ shown in Fig. \ref{Fig1}(a). To be
specific, the BS (with properly defined phase shifts)
realizes transformation%
\begin{equation}
\frac{|1\rangle+i|2\rangle}{\sqrt{2}}\rightarrow|1\rangle,\qquad
\frac{|1\rangle-i|2\rangle}{\sqrt{2}}\rightarrow|2\rangle,\label{eq bs}%
\end{equation}
The two unitary transformations on the polarization states are designed to be%
\begin{equation}
U_{B}=\left(
\begin{array}
[c]{cc}%
\frac{\sqrt{3}}{2} & \frac{1}{2}\\
\frac{1}{2} & -\frac{\sqrt{3}}{2}%
\end{array}
\right)  ,\qquad U_{C}=\left(
\begin{array}
[c]{cc}%
\frac{1}{2} & \frac{\sqrt{3}}{2}\\
\frac{\sqrt{3}}{2} & -\frac{1}{2}%
\end{array}
\right)  ,\label{eq 4}%
\end{equation}
$\allowbreak\allowbreak$which realize transformation%
\begin{align}
U_{B}|B+\rangle &  =U_{C}|C+\rangle=|\text{H}\rangle,\nonumber\\
U_{B}|B-\rangle &  =U_{C}|C-\rangle=|\text{V}\rangle.\label{eq 5}%
\end{align}
Hence the unitary transformation shown in Eq. (\ref{eq 14}) for this case is realized.

According to Ref. \cite{eng}, $U_{B}$, and $U_{C}$ could be realized by only
half-wave plates (HWP). The HWP, with its major axis at an angle $\theta$ to
the vertical direction, is accounted for the unitary operator (up to an
overall phase factor)%
\begin{equation}
\text{HWP}\left(  \theta\right)  =\left(
\begin{array}
[c]{cc}%
\cos2\theta & \sin2\theta\\
\sin2\theta & -\cos2\theta
\end{array}
\right)  .\label{eq hwp}%
\end{equation}
Therefore we have%
\begin{equation}
U_{B}=\text{HWP}\left(  \theta_{B}\right)  ,\qquad U_{C}=\text{HWP}\left(
\theta_{C}\right)  ,\label{eq angle}%
\end{equation}
with%
\begin{equation}
\theta_{B}=\frac{1}{2}\arccos\frac{\sqrt{3}}{2}=15%
{{}^\circ}%
,\qquad\theta_{C}=\frac{1}{2}\arccos\frac{1}{2}=30%
{{}^\circ}%
.
\end{equation}

The schematic drawing of the implementation of POVM $\{\mathcal{E}_{B\pm
},\mathcal{E}_{C\pm}\}$ is shown in Fig. \ref{Fig2}(b). The POVM
$\{\mathcal{E}_{A\pm},\mathcal{E}_{B\pm}\}$ ($\{\mathcal{E}_{C\pm}%
,\mathcal{E}_{A\pm}\}$) could be implemented by simply removing HWP$\left(
\theta_{C}\right)  $ [HWP$\left(  \theta_{B}\right)  $] in Fig. \ref{Fig2}(b),
with the detectors that previously corresponds to $\mathcal{E}_{C\pm}$
($\mathcal{E}_{B\pm}$) now corresponding to $\mathcal{E}_{A\pm}$.

\begin{table}[ptb]
\caption{The experimental data counted in $10$ seconds. For each
POVM, \textquotedblleft\textit{1-fold counts}\textquotedblright\
shows the counts that only one operator yields the answer
\textit{yes} with coincidence with the detector of photon $2$,
while \textquotedblleft\textit{2-fold counts}\textquotedblright\
means that two operators simultaneously yield answer \textit{yes}
with coincidence with the detector of photon $2$. The data in
\textquotedblleft\textit{2-fold counts}\textquotedblright\ has
been scaled according to the carefully measured efficiency of our
single photon detectors and is hence comparable with data in
\textquotedblleft\textit{1-fold counts}\textquotedblright. In our
experiments, the 3- and 4-fold coincidence counts, corresponding
to that more than two operators yield the answer \textit{yes},
turn out to be virtually zero in $10$ seconds.}
\label{exp-data}%
\begin{ruledtabular}%
\begin{tabular}{cccccccccc}
\multicolumn{2}{c}{$\{\mu _{\pm },\nu _{\pm }\}$} & \multicolumn{4}{c}{%
\textit{1-fold counts}} & \multicolumn{4}{c}{\textit{2-fold counts}} \\
\cline{1-2}\cline{3-6}\cline{7-10}
$\mu $ & $\nu $ & $\mu _{+}$ & $\mu _{-}$ & $\nu _{+}$ & $\nu _{-}$ & $%
\begin{array}{c}
\mu _{+} \\
\nu _{+}%
\end{array}%
$ & $%
\begin{array}{c}
\mu _{+} \\
\nu _{-}%
\end{array}%
$ & $%
\begin{array}{c}
\mu _{-} \\
\nu _{+}%
\end{array}%
$ & $%
\begin{array}{c}
\mu _{-} \\
\nu _{-}%
\end{array}%
$ \\ \hline
$\mathcal{E}_{A}$ & $\mathcal{E}_{B}$ & $14718$ & $10474$ & $13156$ & $12587$
& $34$ & $69$ & $38$ & $40$ \\
$\mathcal{E}_{B}$ & $\mathcal{E}_{C}$ & $10660$ & $14781$ & $11902$ & $12103$
& $95$ & $47$ & $85$ & $63$ \\
$\mathcal{E}_{C}$ & $\mathcal{E}_{A}$ & $12883$ & $10586$ & $13764$ & $10940$
& $128$ & $39$ & $18$ & $24$%
\end{tabular}%
\end{ruledtabular}
\end{table}

Here we shall note that although our experiment and that in Ref. \cite{r5} are
both based on single photons, there are substantial differences. In theory,
our experiment tests the KS theorem for a single two-level system. The path
degrees of freedom are used as ancilla, which according to Ref. \cite{cabello}
should be regarded as part of the measurement apparatus, which can be
considered to arise from the beam splitter-induced \textquotedblleft
interference\textquotedblright\ between the photon to be measured and the
vacuum. While in Ref. \cite{r5} the path degrees of freedom are part of the
system to be tested. In practice, our experiment does not demand the full MZ
setup, and is irrelevant to relative phases between paths. Hence our
experiment is much simpler and more convenient.

In experiments, we generate two photons (labelled by $1$ and $2$)
in the maximally entangled state, with a visibility of about
$82\%$, $|\Psi^{-}\rangle_{12}=\frac{1}{\sqrt{2}}\left(
|\text{H}\rangle_{1}|\text{V}\rangle_{2}-|\text{V}\rangle_{1}|\text{H}%
\rangle_{2}\right)  $ by type-II spontaneous parametric down
conversion (SPDC) from a pump pulse passing through a beta-barium
borate (BBO) crystal. The UV laser with a central wavelength of
$394$nm has a pulse duration of $80$fs, a repetition rate of
$80$MHz, and an average power of $110$mw. By tracing out photon
$2$, photon $1$ is left in a maximal mixed state described by
density matrix
$\rho_{1}=\frac{1}{2}\left(  |\text{H}\rangle\langle\text{H}|+|\text{V}%
\rangle\langle\text{V}|\right).$ The three POVMs,
$\{\mathcal{E}_{A\pm },\mathcal{E}_{B\pm}\}$,
$\{\mathcal{E}_{B\pm},\mathcal{E}_{C\pm}\} $, and
$\{\mathcal{E}_{C\pm},\mathcal{E}_{A\pm}\}$, are performed on
photon $1$ at state $\rho_{1}$.

The experimental data contained in Table \ref{exp-data} shows the
number of the events in which \textit{one and only one} operator
yields the answer \textit{yes}, and the number of those in which
more than one operator simultaneously yield the answer
\textit{yes}. The collection and detection efficiencies of the
four port are $\sim5\%$ in our experiments. Through careful
calculation, the 2-fold coincidence has been scaled to be
comparable to 1-fold data and the experimental results coincide
with a very high precision ($\sim99\%$) with Eqs. (\ref{eq 2}),
which therefore experimentally excludes the existence of a
non-contextual hidden-variable theory even for a single qubit.
From the experimental point of view, these 2-fold counts are due
to the imperfect entangled photon source. In our experiments,
because of the probablistic feature of pair creation in SPDC,
there will be a small probability that more than one pair is
generated. The additional pair(s) will give some 2-fold counts
(about $150$), which is of the same order of the counts observed
in our experiments as presented in Table \ref{exp-data}.

In conclusion, we propose an experimental scheme for the implementation of
arbitrary positive operator-valued measures on the polarization states of
single photons using linear optical devices. This scheme may have various
applications in quantum information processing. As a demonstration, we present
the experimental test of the KS theorem for a single qubit using POVM. Our
experiment verifies with very good precision that even a single qubit could
not be described by NCHV theory.

We thank Z.B. Chen for fruitful discussions and valuable
suggestions. This work was supported by the Nature Science
Foundation of China, the Chinese Academy of Sciences, and the
National Fundamental Research Program (under Grant No.
2001CB309300).

\end{document}